\documentclass[10pt,conference]{IEEEtran}

\usepackage{cite}
\usepackage{amsmath,amsfonts}
\usepackage{algorithmic}
\usepackage{graphicx}
\usepackage{textcomp}
\usepackage{footmisc}
\usepackage{xcolor}
\usepackage{multirow}
\usepackage{siunitx}
\usepackage{balance}
\usepackage{graphics}
\usepackage[T1]{fontenc}
\usepackage{txfonts}
\usepackage{mathptmx}
\usepackage{booktabs}
\PassOptionsToPackage{hyphens}{url}
\usepackage{microtype}
\usepackage{ccicons}
\usepackage{framed}
\usepackage[table]{xcolor}
\usepackage{amsmath,amssymb,amsfonts}
\usepackage{algorithmic}
\usepackage{graphicx}
\usepackage{textcomp}
\usepackage[ampersand]{easylist}
\usepackage{amssymb}
\usepackage{multicol}
\usepackage{multirow}
\usepackage{tabularx}
\usepackage{rotating}
\usepackage[roman]{parnotes}
\usepackage[many]{tcolorbox}
\usepackage{soul}
\usepackage[table]{xcolor}
\usepackage{comment}
\usepackage{booktabs}
\usepackage{colortbl}
\usepackage{verbatim}
\usepackage{array}
\usepackage[inline]{enumitem}
\usepackage[export]{adjustbox}
\usepackage{pifont}
\usepackage{mathtools}
\usepackage{hyphenat}

\begin{document}

\title{Reverse Engineering User Stories from Code using Large Language Models}

\author{\IEEEauthorblockN{Mohamed Ouf}
\IEEEauthorblockA{\textit{Queen's University} \\
Kingston, Canada \\
24blr2@queensu.ca}
\and
\IEEEauthorblockN{Haoyu Li}
\IEEEauthorblockA{\textit{Queen's University} \\
Kingston, Canada \\
17hl66@queensu.ca}
\and
\IEEEauthorblockN{Michael Zhang}
\IEEEauthorblockA{\textit{Queen's University} \\
Kingston, Canada \\
18jz111@queensu.ca}
\and
\IEEEauthorblockN{Mariam Guizani}
\IEEEauthorblockA{\textit{Queen's University} \\
Kingston, Canada \\
mariam.guizani@queensu.ca}}

\maketitle

\begin{abstract}

User stories are essential in agile development, yet often missing or outdated in legacy and poorly documented systems. We investigate whether large language models (LLMs) can automatically recover user stories directly from source code and how prompt design impacts output quality. Using 1,750 annotated C++ snippets of varying complexity, we evaluate five state-of-the-art LLMs across six prompting strategies. Results show that all models achieve, on average, an F1 score of 0.8 for code up to 200 NLOC. Our findings show that a single illustrative example enables the smallest model (8B) to match the performance of a much larger 70B model. In contrast, structured reasoning via Chain-of-Thought offers only marginal gains, primarily for larger models.

%User stories play a central role in agile development by guiding teams to deliver features focused explicitly on user value. However, these stories are often missing or outdated, especially in legacy or poorly documented systems. In this paper, we investigate the potential of Large Language Models (LLMs) to automatically recover user stories directly from source code, exploring how prompt design influences the quality of generated stories. We systematically evaluate five state-of-the-art LLMs across six prompting strategies using a diverse dataset comprising 1,750 annotated C++ code snippets with varying complexities. Evaluating model outputs against human-generated user stories using BERTScore, we find all models achieve adequate or better performance for code snippets under 200 lines of code. Notably, providing a single illustrative example significantly improves the smallest model (8 billion parameters), allowing it to match the performance of a much larger 70-billion-parameter baseline. In contrast, incorporating structured reasoning prompts (Structured Chain-of-Thought) yields only minor benefits, and primarily for larger models. 
%Our findings emphasize that careful prompt engineering is more effective and efficient than merely scaling model size, suggesting a practical, low-cost approach for automating backlog reconstruction and improving agile processes.
\end{abstract}


\section{Introduction}
\label{sec:intro}

 User stories are a cornerstone of agile development, capturing stakeholder needs and guiding feature delivery with a focus on user value. Typically structured as \textit{``As a [user role], I want [goal] so that [benefit]''}, these concise narratives align development activities with business goals and user expectations \cite{lucassen2015forging, ramesh2010agile}. Recent research has explored automating user story generation from textual artifacts during requirements elicitation. For instance, Rahman and Zhu used GPT-4 to distill requirement documents into user stories \cite{rahman2024automated}, while Veitia et al. applied NLP techniques to extract stories from software issue records \cite{chinnaswamy2024user}. These methods reduce reliance on manual effort but require existing textual specifications.

In contrast, source code remains the most persistent and widely available artifact, often outliving design notes, meeting records, and other documentation especially in legacy systems or after team turnover. This makes code a promising source for automated recovery of user stories, particularly in contexts where traditional documentation is sparse or outdated. Automatically generating user stories from code has both technical and managerial value. For example, agile teams can use code-derived stories during retrospectives to compare intended and implemented functionality. New team members can use them to understand the goals behind features, improving onboarding.

Unlike code summaries, which describe how a system functions (e.g., \textit{``This module implements Dijkstra’s algorithm to compute shortest paths in a graph''}), user stories emphasize the underlying user need and value (e.g., \textit{``As a logistics manager, I want to find the most efficient delivery routes so that I can reduce transportation costs and meet delivery deadlines''}). This shift from implementation to intent makes user stories more accessible to non-technical stakeholders—product managers, clients, and end-users alike—and enables more intuitive progress tracking. Open-source software (OSS) communities stand to benefit significantly from this approach due to issues such as fragmented documentation, high contributor turnover, and barriers to entry \cite{guizani2021long, guizani2022perceptions, steinmacher2015systematic}. For newcomers, clearly articulated user stories can foster understanding and inclusion, potentially improving retention and engagement. In both OSS and industry settings, reverse engineering user stories can help navigate undocumented legacy systems.

\begin{figure*}[h]
\centering
\includegraphics[width=15cm, height=3cm]{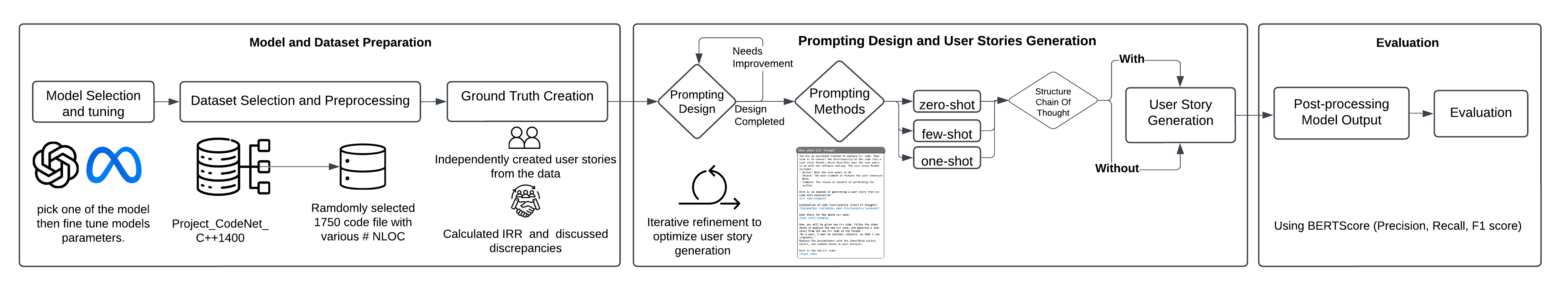}
\caption{Overview of our Approach.}
\label{fig:oneshot_prompt_with_cot}
\end{figure*} 

Despite its broad potential, the automatic recovery of user stories directly from source code remains largely unexplored. To our knowledge, this is the first work to investigate this problem. We explore its feasibility and provide the first empirical foundation for reverse engineering user stories across varying levels of code complexity.

\vspace{0.2em}
\noindent
This paper makes the following contributions:
\begin{itemize}
    \item We present the first study to evaluate the automatic generation of user stories from code.
    \item We release a dataset of 1,750 C++ code–story pairs.
    \item We empirically analyze how model size and prompt design, including structured reasoning (SCoT) and few-shot techniques, impact generation quality.
\end{itemize}

\vspace{0.2em}
\noindent
We address two research questions:
\begin{itemize}
    \item \textbf{RQ1:} \emph{Can large language models generate user stories from code? How do code complexity and model architecture affect story quality?}
    \item \textbf{RQ2:} \emph{To what extent do prompting techniques (e.g., SCoT, k-shot) improve user story generation?}
\end{itemize}

\section{Related Work}
\label{sec:related}

Existing research on user stories has primarily focused on extracting them from textual artifacts, such as test cases \cite{genaid2012connecting,fischbach2020specmate}, diagrams \cite{elallaoui2015automatic,elallaoui2018automatic}, and use case scenarios \cite{gilson2020generating}. Raharjana et al. \cite{raharjana2021user} highlight two major limitations in this space: limited coverage of diverse software artifacts and a lack of benchmark datasets tailored to user-story–related tasks. Significant efforts have explored generating user stories from textual sources. Techniques have been applied to requirements documents \cite{rahman2024automated,nistala2022towards}, issue records \cite{chinnaswamy2024user,colavito2024leveraging}, stakeholder interviews \cite{abed2024ai,marques2022enriching}, and design specifications \cite{gilson2020generating}. For instance, Rahman and Zhu \cite{rahman2024automated} introduced ``GeneUS'', a GPT-4-powered tool that transforms requirement documents into structured user stories, while Abed et al. \cite{abed2024ai} investigated the use of AI for generating user stories from stakeholder interviews and evaluated the results using the Quality User Stories (QUS) framework. However, these approaches all rely on the presence of existing textual specifications and do not extend to generating user stories directly from code. Large Language Models (LLMs) have increasingly been adopted for various software engineering tasks, including code refactoring \cite{marques2024using,tikayat2023agile}, commit message generation \cite{lee2024github}, debugging \cite{macneil2024decoding}, and requirements elicitation \cite{marques2024using}. As demonstrated by Wong et al. \cite{wong2023natural}, transformer-based LLMs trained on large code corpora can capture both syntactic and semantic program structures—enabling new capabilities in downstream tasks such as automated code summarization.

The closest body of work to ours is code summarization. This field has evolved from rule-based approaches \cite{kim2012identifying,zhang2022survey}, to neural models \cite{bansal2021project,leclair2021ensemble}, and more recently, to LLM-based techniques \cite{sun2024source,yun2024project}. Zhu et al. \cite{zhu2024effectiveness} show that fine-tuning LLM parameters improves summarization performance. However, while code summarization aims to describe what the code does, primarily for technical audiences, our goal is to surface why the code exists, capturing user-facing intent and benefits. This distinction makes user stories more suitable for diverse stakeholders, including non-developers and newcomers in open-source or distributed development settings. Despite the growth in both user story generation from text and LLM-based code summarization, no prior work has explored generating user stories directly from source code. This paper addresses this gap.

\section{Methodology}

Figure~\ref{fig:oneshot_prompt_with_cot} illustrates our approach for reverse engineering user stories from source code using LLMs.

\subsection{Dataset Selection and Ground-Truth Development}
\label{sec:data}

C++ ranks 6th on GitHub with 10\% of public contributions \cite{octoverse2022} and remains in the TIOBE top 3 throughout 2024 \cite{tiobe2024}. Its prevalence makes it well-suited for studying reverse-engineered user stories.

We use the \texttt{Project\_CodeNet\_C++1400} subset of IBM CodeNet, which contains approximately 420,000 accepted C++ solutions to 1400 competitive‐programming tasks \cite{puri2021codenet, ibm_codenet_2021}. 

%Competitive problems target real‐world kernels (routing, scheduling, resource allocation), and an analysis shows that token and function‐length distributions in contest code mirror those in large open‐source C/C++ projects \cite{hatton2017length}.

%\textbf{Stratified sampling, real-world complexity, and context length.}
%To choose realistic size bands, we first cloned the \emph{top 200 C++ GitHub
%repositories by stars} (snapshot March 2025) and computed
%non-commenting lines of code (NLOC) for 950 classes and 15 400functions. Functions typically contain about 55 NLOC, while classes range from 100 to 200 NLOC. Guided by this, we defined 35 equal-width strata (1–10, 11–20, …, 341–350 NLOC) andrandomly sampled 50 files per stratum from CodeNet, yielding 1 750 codesnippets.  The smallest prompts (zero-shot on a 1–10 NLOC file) occupy 350 input tokens, whereas the largest configuration(few-shot + SCoT on a 341–350 NLOC file) approaches 3 500 tokens.Because mainstream LLMs exhibit the ``lost in the middle'' degradationwhen important signals lie deep inside long contexts \cite{liu2023lost},our span of sizes simultaneously tests logical complexity and context-window robustness.

To define realistic size bands, we cloned the top 200 C++ GitHub repositories by stars (as of March 2025) and measured non-commenting lines of code (NLOC) across 950 classes and 15400 functions. Functions averaged 55 NLOC, while classes ranged from 100 to 200. Based on this, we created 35 equal-length strata (1–10, 11–20, …, 341–350 NLOC) and randomly sampled 50 CodeNet files per stratum, yielding 1,750 snippets. We highlight the 100 to 200 NLOC as our range of interest, the grayed range in our figures in the findings section. Prompt lengths ranged from 350 input tokens (zero-shot, 1–10 NLOC) to 3500 tokens (few-shot + SCoT, 341–350 NLOC). This range tests both logical complexity and LLM robustness to the ``lost in the middle'' effect \cite{liu2023lost}.

%\textbf{Ground-truth user stories.}
Three C++ proficient researchers independently authored user stories for
each code snippet, then reconciled differences; inter-rater reliability
measured via Cohen’s $\kappa$ was 0.86, indicating almost perfect
agreement \cite{cohen1960coefficient}.  Our dataset mapping 1,750 C++ code snippets of varying complexity to user stories is available in the supplemental material\cite{suppdoc}.

\subsection{Model Selection, Characteristics \& Configuration}
\label{sec:model_setup}
Our study evaluates five large language models chosen to capture the effect of \emph{parameter count}, \emph{architecture}, and \emph{economic cost} shown in Table \ref{tab:model_costs} on the automatic generation of user stories from code.

%\textbf{Parameter‐controlled family (open source).}
We use three instruction-tuned variants of \textsc{LLAMA 3.1}—8B, 70B, and 405B parameters. These models share the same transformer architecture and training data, but differ by two orders of magnitude in scale. Because the 8B model can run locally on a single GPU, the 70 B requires a modest cloud instance, and the 405 B needs multi-GPU servers, these three models isolate the ``more parameters $\Rightarrow$ better quality'' hypothesis without confounding architectural changes.

%\textbf{State-of-the-art reasoning model.}
We include DeepSeek-\textsc{R1} to investigate whether architectural
optimizations for step-by-step reasoning translate into higher-quality
user stories. DeepSeek-\textsc{R1} augments the vanilla transformer with explicit reasoning controllers and built-in structured chain-of-thought (SCoT). 

%\textbf{Closed-source production baseline.}
OpenAI’s \textsc{GPT-4o-mini} represents a widely-used commercial LLM.
We test it ``as-is'' and with lightweight fine-tuning (70 in-context
examples) to gauge the marginal benefits of task-specific adaptation.

\begin{table}[h]
\centering
\caption{Model cost per-$1\mathrm M$-token (USD).}
\label{tab:model_costs}
\resizebox{1\columnwidth}{!}{%
\begin{tabular}{lcc|lcc}
\toprule
Model & Input & Output & Model & Input & Output \\
\midrule
\textsc{Llama 3.1-8B}   & 0.05\$  & 0.25\$~\cite{meta8bprice} &
\textsc{Llama 3.1-70B}  & 0.65\$  & 2.75\$~\cite{meta8bprice} \\

\textsc{Llama 3.1-405B} & 9.50\$  & 9.50\$~\cite{meta8bprice} &
DeepSeek-\textsc{R1}    & 0.55\$  & 2.19\$~\cite{deepseekprice} \\

\textsc{GPT-4o-mini}    & 0.80\$  & 3.20\$~\cite{openaiapi}    &
\textsc{O1}             & 10.00\$ & 40.00\$~\cite{openaiapi}        \\
\bottomrule
\end{tabular}%
}
\end{table}

%\textbf{Token-level economics.}

Table~\ref{tab:model_costs} lists the pay-as-you-go prices (USD per
$10^{6}$ tokens) advertised at the time of writing.  While
input tokens, costs rise steeply with scale and proprietary licensing.
The new GPT-O1 reasoning model is \textasciitilde\!\,200$\times$ more expensive for inputs and \textasciitilde\!\,160$\times$ for outputs, and even standard GPT-4o is three to five times costlier than DeepSeek-\textsc{R1}.
%
%\textbf{Completion length.}
Reasoning-centric models are more verbose. In our experimentation,  DeepSeek-\textsc{R1} averaged \SI{620} output tokens (occasionally exceeding~\SI{2500}), roughly a \SI{5}{\times} increase over the 100 to 130 token range observed for \textsc{LLAMA-70B},
and \textsc{GPT-4o mini}.

\subsection{Prompt Design}
\label{prompt-design}

We designed prompts through iterative refinement by two researchers who met on a weekly basis to discuss and refine the prompts. 
This process resulted in six templates combining three techniques zero-shot, one-shot, few-shot, with and without Structured Chain of Thought (SCoT) reasoning. All prompt variants are available in Supplemental Material \cite{suppdoc}. Unlike standard Chain-of-Thought (CoT), which produces free-form reasoning in natural language \cite{fujita2024llm}, Structured Chain-of-Thought (SCoT) imposes programmatic structure by decomposing reasoning into sequence, branch, and loop primitives \cite{li2023structured}. This alignment with code structure (e.g., control flow) makes SCoT suitable for reverse engineering user stories.

%\begin{enumerate}
%    \item Code execution paths map directly to user workflows
 %   \item Control structures (e.g., loops) imply iterative user actions
%    \item Conditional branches represent decision points in user goals
%\end{enumerate}
%\cite{li2023structured}, SCoT provides scaffolding to infer user intent from implementation patterns.

%- \textit{Zero-shot}: No examples, testing base capability to understand the task and do it  
%- \textit{One-shot}: Single example demonstrating input-output mapping  
%- \textit{Few-shot}: two examples covering diverse complexity bands.

We added directives (e.g., \textit{``Output user stories in 'As a...' format''}) at the beginning of the prompt to combat the ``lost in the middle'' effect \cite{liu2023lost}, where LLMs might forget central content in long contexts. This optimization improved instruction adherence by 22\% in pilot tests.

%\begin{figure}[h]
%\centering
%\includegraphics[width=8.5cm, height=9cm]{figures/Prompt_example.png}
%\caption{One-shot prompt with Chain of Thought (CoT) Reasoning.}
%\label{fig:oneshot_prompt_with_cot}
%\end{figure}

\subsection{Evaluation of Semantic Similarity Metrics}
\label{sub:semantic_metrics}
\begin{table}[!t]
\centering
\caption{Average similarity scores.}
\small
\setlength{\tabcolsep}{4pt}
\renewcommand{\arraystretch}{0.8}

\resizebox{0.85\linewidth}{!}{%
  \begin{tabular}{@{} l l c c c @{}}
    \toprule
    Metric            & Model                      & Twin‐Minimal & Paraphrase‐50 & Different‐Meaning \\
    \midrule
    \multirow{3}{*}{BERTScore‐P (F1)} 
                      & bert‐base‐uncased           & 97           & 85            & 60                 \\
                      & roberta‐large               & 99           & 96            & 81                 \\
                      & microsoft/deberta‐large     & 97           & 90            & 84                 \\
    \midrule
    \multirow{2}{*}{BLEU‐P (F1)} 
                      & default                     & 90           & 65            & 56                 \\
                      & smoothing                   & 90           & 65            & 56                 \\
    \midrule
    \multirow{2}{*}{ROUGE‐L (F1)} 
                      & default                     & 89           & 55            & 50                 \\
                      & no‐stem                     & 89           & 55            & 50                 \\
    \bottomrule
  \end{tabular}%
}

\label{tab:similarity_comparison}
\end{table}

To identify the most suitable metric for comparing LLM-generated user stories against ground-truth references, we conducted an experiment using 60 curated user-story pairs divided into three categories. The Twin-Minimal category included 20 pairs with only minor lexical changes that preserved identical meaning (e.g., ``sum two large numbers'' $\rightarrow$ ``add two significant numbers''). The Paraphrase-50 category consisted of 20 pairs with approximately 50\% lexical variation while maintaining semantic equivalence (e.g., ``merge tasks into streamlined schedule'' $\rightarrow$ ``organize tasks into clear schedule''). The Different-Meaning category featured 20 pairs exhibiting fundamental semantic divergence (e.g., ``encrypted storage for document security'' $\rightarrow$ ``compression for storage reduction''). Each pair was evaluated using multiple metrics: three BERTScore variants (bert-base-uncased, roberta-large, and microsoft/deberta-large), ROUGE (both default and no\_stem), and BLEU (with and without smoothing). Table~\ref{tab:similarity_comparison} summarizes the average precision scores in these categories.

%To identify the metric for comparing LLM-generated user stories against ground truth references, we experimented using 60 selected user-story pairs divided into three distinct categories:

%We defined three evaluation categories to test metric sensitivity:
%\begin{enumerate}
%    \item \textbf{Twin-Minimal (20 pairs):} Minor lexical changes preserving identical meaning (e.g., \emph{"sum two large numbers"} $\rightarrow$ \emph{"add two significant numbers"})
    
%    \item \textbf{Paraphrase-50 (20 pairs):} $\sim$50\% lexical variation with preserved semantics (e.g., \emph{"merge tasks into streamlined schedule"} $\rightarrow$ \emph{"organize tasks into clear schedule"})
    
%    \item \textbf{Different-Meaning (20 pairs):} Fundamental semantic divergence (e.g., \emph{"encrypted storage for document security"} $\rightarrow$ \emph{"compression for storage reduction"})
%\end{enumerate}

%We evaluated each pair using multiple metrics: three BERTScore variants (\texttt{bert-base-uncased}, \texttt{roberta-large}, \texttt{microsoft/deberta - large}), ROUGE (default and no\_stem), and BLEU (with/without smoothing). Table \ref{tab:similarity_comparison} summarizes average precision scores across categories.

The result showed that BLEU and ROUGE have marginal differentiation between Paraphrase-50 (65 and 55 respectively) and Different-Meaning pairs (56 and 50). Both BLEU and ROUGE fail to distinguish between outputs that preserve the intended meaning (semantic equivalence) and those that do not. Among BERTScore variants, \texttt{roberta-large} and \texttt{microsoft/ deberta-large} has limited discrimination for meaning-altered pairs (Different-Meaning scores: 81 and 84). Only \texttt{bert-base-uncased} demonstrated a good scoring gradient: near-perfect for Twin-Minimal (97), lower for Paraphrase-50 (85), and distinct for Different-Meaning (60).

Therefore, we adopt \texttt{bert-base-uncased} and interpret its F1 scores as follows: %based on the min-max to \textit{Paraphrase‐50}:

\begin{itemize}

  \item Faithful generation (\(\geq 0.9\))
  \item Adequate paraphrases (0.66 – 0.9)
  \item Meaning-divergent generation (\(\leq 0.65\))
\end{itemize}

\subsection{Experimental Setup}
We investigate three core dimensions influencing user story generation from source code. First, we evaluate model architecture effects by comparing five large language models representing distinct design paradigms: parameter-scaled variants of the LLAMA 3.1 family (8B, 70B, and 405B parameters) to isolate scale-dependent capabilities; the reasoning-optimized DeepSeek-R1 architecture; and OpenAI's production-grade GPT-4o-mini both in its base configuration and a fine-tuned alternative. All models were deployed through the Replicate platform \cite{replicate2024} to ensure consistent execution environments. 

%, with notable cost considerations for DeepSeek-R1 (\$10 per million tokens) due to frequent unavailability of its native API during our study period.

Second, we measure the impact of Structured Chain-of-Thought (SCoT) reasoning augmentation across all models. Each model was evaluated in paired configurations: with standard prompting and with SCoT-enhanced prompting. Third, we conduct focused prompt engineering on the smallest model (LLaMA-8B) to determine if exemplar-based techniques can compensate for parameter limitations. This investigation employs six prompt variants: three strategies (zero-shot, one-shot, few-shot) each tested with and without SCoT augmentation to isolate improvement potential from both knowledge injection (via examples) and structured reasoning.

All experiments share consistent decoding parameters: temperature=0, minimum output length=50 tokens, and repetition penalty=0.2. Evaluations are stratified across 35 code complexity bands (1-10 to 341-350 non-comment lines of code). 

\section{Findings}
\subsection{RQ1: Can LLMs generate user stories from code? And how do different code complexity levels and LLMs architectures factor in the generated user stories?}

\label{sub:rq1_findings}
\begin{figure}[h]
  \centering
    \caption{BERTScore F1 across all NLOC range across all different models using zero-shot prompting without SCOT.}
  \includegraphics[width=\columnwidth]{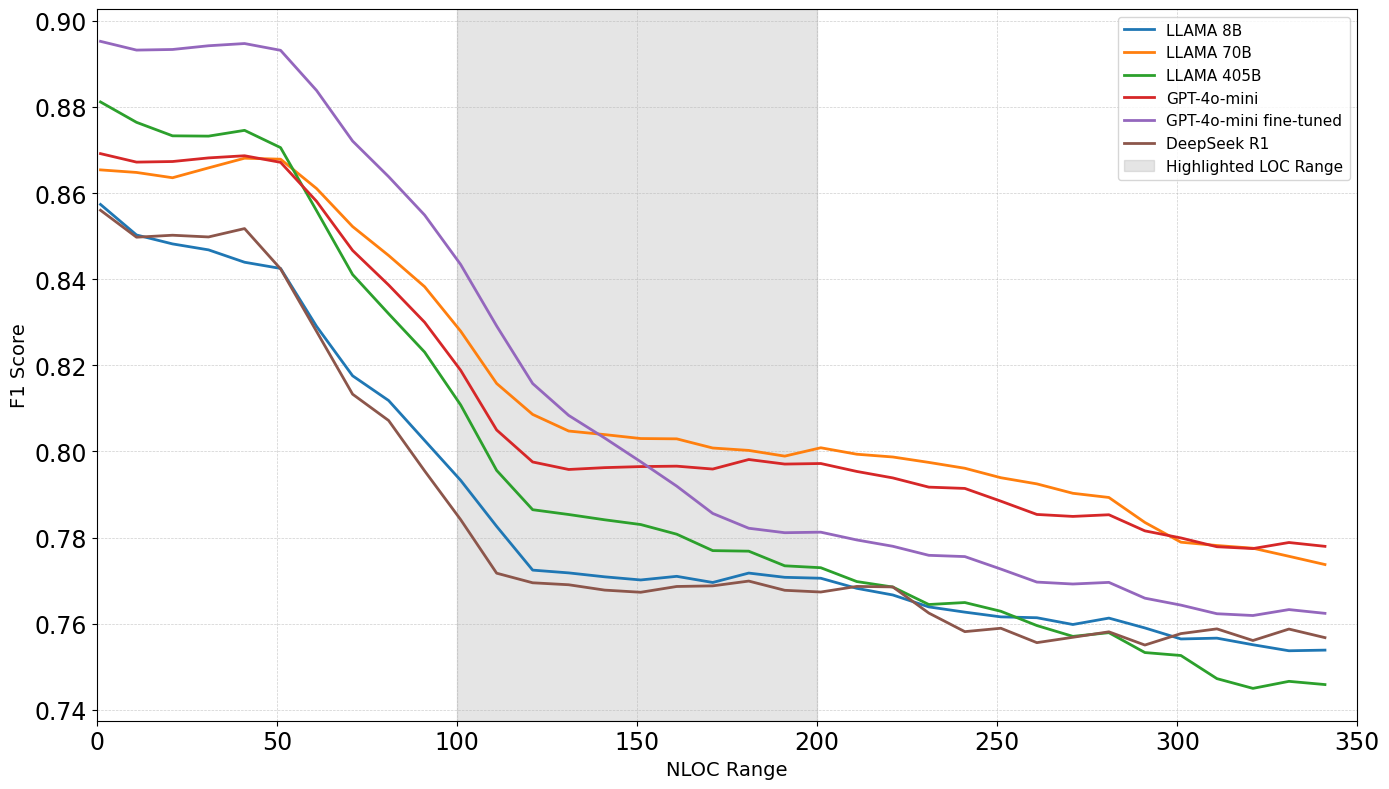}
  \label{fig:f1_all_models}
\end{figure}

\subsubsection{Overall performance}
Our findings show that overall, LLMs can generate user stories from code.
However, across all six models, the F1 score decreases as NLOC increases (See Fig.~\ref{fig:f1_all_models} and Table \ref{tab:averaged_scores}). This highlights that despite overall adequate performance scaling the generation of user stories from code, our future research focus, can be challenging. For code with less than 100 NLOC, every model achieves an F1 score of $0.78$ or above. None of the models’ average F1 scores were below $0.83$ as shown in Table \ref{tab:averaged_scores}, which is within the high end of the \emph{Adequate-Paraphrasing} threshold (0.66 - 0.9), where the generated user story exhibits different lexical variations while maintaining its overall meaning. For code within the 100 to 200 NLOC range, which represents functionality complexity in practice, F1 scores range from $0.77$ to $0.84$, maintaining \emph{Adequate Paraphrasing} quality.

Even when exceeding 200 NLOC, none of the models falls below the $0.74$ F1 score. This highlights that no model crosses the \emph{Meaning-divergent} boundary (\(\leq 0.65 \)), where the semantic meaning of the user story is compromised. The decreasing F1 in the 200+ NLOC range could be attributed to the ``lost-in-the-middle'' effect, where tokens in the middle of long prompts have less attention \cite{liu2023lost}.

\subsubsection{Model–specific trends}

As illustrated in Figure~\ref{fig:f1_all_models}, the LLAMA-8B model consistently underperforms relative to its larger counterparts, the LLAMA-70B and LLAMA-405B models. Notably, the LLAMA-8B model shows a marked decline in performance, particularly at lower levels of code complexity, where it consistently achieves the lowest F1 scores up to 255 NLOC. Beyond this threshold, the performance of the LLAMA-405B model falls below that of the LLAMA-8B, underscoring the challenges smaller models face in generating user stories, even for relatively simple code samples.

In contrast, the performance trajectories of the LLAMA-70B and LLAMA-405B are similar up to 100 NLOC, at which point the difference in performance increases. LLAMA-70B model consistently outperforms the LLAMA-405B across increasingly complex code samples (> 55 NLOC), and this performance gap continues to expand as complexity increases. This highlights that up to a certain number of parameters, higher parameter count does not necessarily result in a performance increase.

Given that the 70B model costs significantly less than the 405B variant (Table~\ref{tab:model_costs}), it represents the most cost-effective open-source option. 

%Given the substantial cost difference between the 70B and 405B models (Table~\ref{tab:model_costs}), the LLAMA 70B model emerges as the most cost-effective and efficient open-source option for user story generation especially when NLOC increases.

%Within the parameter-controlled LLAMA family, model size demonstrates performance benefits. As shown in Figure~\ref{fig:f1_all_models}, the LLAMA 8B model consistently underperforms compared to its larger counterparts, the LLAMA 70B and LLAMA 405B models. The LLAMA 8B model shows the lowest performance for lower code complexity compared to LLAMA 70B and LLAMA 405B. It consistently achieves the lowest F1 scores up to 255 NLOC, after which the LLAMA 405B performance drops below the LLAMA 8B model. Suggesting that smaller models struggle in generating user stories, even for less complex code samples. The LLAMA 70B and LLAMA 405B curves show similar performance up to 50 NLOC, after which they diverge, with the LLAMA 70B model consistently outperforming the LLAMA 405B. This performance gap widens across complexity bands.

Despite its larger parameter count, DeepSeek-R1 performs similarly to LLAMA-8B across all code complexity levels as shown in Table \ref{tab:averaged_scores}. This suggests that parameter count is not always equated with better performance. 

This might be due to the fact that the reasoning architecture of DeepSeek-R1 was trained mainly on code and mathematics tasks\cite{deepseek2024r1}, whereas user-story generation requires capturing both code functionality and the human goals and intentions behind it.

Our findings show that light fine-tuning of GPT-4o-mini produces performance improvement for small to medium NLOC ranges (see Figure \ref{fig:f1_all_models}). It is worth noting that fine-tuning does have the added benefit of reducing output token usage by 80\% which can be particularly useful when inference is run at scale 

%since the adapted model generates concise user stories without explanatory text.

%produces minimal performance improvements in the lower half of NLOC range, this trend shifts in the upper half of NLOC (see Figure\ref{fig:f1_all_models}). 

%Despite these variations, all models maintain F1 scores within the adequate paraphrasing range 
%across all complexity levels, ensuring correct user story generation.

\begin{center}
\fbox{\parbox{0.94\linewidth}{%
\textbf{RQ1 Answer.} Overall, LLMs can adequately generate user stories from source code, averaging F1 scores of 0.8 for code up to 200 NLOC.
%achieving F1 scores above 0.74 across all complexity levels. 
Up to a certain model size, increasing the number of parameters has diminishing returns. Reasoning models do not outperform standard models, while light fine-tuning a smaller model improves performance for small to medium NLOC ranges and reduces output size by 80\%.
}}
\end{center}

\begin{table*}[ht]
\caption{Comparison of Ground Truth vs.\ Generated User Stories with average precision, recall and F1 scores, bold results = increasing from No SCOT to with SCOT.}
\centering
\begin{adjustbox}{max width=0.9\textwidth}
\begin{tabular}{cl|cc|cc|cc|cc|cc}
\toprule
\textbf{NLOC} &  & \multicolumn{2}{c|}{\textbf{LLAMA 8B}} & \multicolumn{2}{c|}{\textbf{LLAMA 70B}} & \multicolumn{2}{c|}{\textbf{LLAMA 405B}} & \multicolumn{2}{c|}{\textbf{GPT 4o-mini}} & \multicolumn{2}{c}{\textbf{GPT 4o-mini Fine-tuned}} \\
\cmidrule(lr){3-4} \cmidrule(lr){5-6} \cmidrule(lr){7-8} \cmidrule(lr){9-10} \cmidrule(lr){11-12}
& & No SCOT & With SCOT & No SCOT & With SCOT & No SCOT & With SCOT & No SCOT & With SCOT & No SCOT & With SCOT \\
\midrule
\multirow{3}{*}{1–100} 
& P  & 0.82 & 0.81 & 0.87 & 0.85 & 0.86 & 0.84 & 0.87 & 0.84 & 0.88 & 0.86 \\
& R  & 0.84 & 0.82 & 0.85 & 0.83 & 0.85 & 0.84 & 0.85 & 0.83 & 0.89 & 0.87 \\
& F1 & 0.83 & 0.81 & 0.86 & 0.84 & 0.86 & 0.84 & 0.86 & 0.84 & 0.88 & 0.87 \\
\midrule
\multirow{3}{*}{101–200} 
& P  & 0.75 & 0.75 & 0.79 & \textbf{0.83} & 0.75 & \textbf{0.78} & 0.78 & 0.78 & 0.74 & 0.73 \\
& R  & 0.79 & 0.77 & 0.82 & \textbf{0.84} & 0.81 & \textbf{0.83} & 0.80 & 0.81 & 0.79 & 0.78 \\
& F1 & 0.77 & 0.77 & 0.80 & \textbf{0.83} & 0.78 & \textbf{0.80} & 0.79 & 0.79 & 0.76 & 0.75 \\
\midrule
\multirow{3}{*}{201-350} 
& P  & 0.72 & 0.72 & 0.76 & \textbf{0.78} & 0.72 & \textbf{0.73} & 0.76 & 0.76 & 0.76 & 0.75 \\
& R  & 0.79 & 0.79 & 0.80 & \textbf{0.83} & 0.78 & \textbf{0.81} & 0.80 & 0.80 & 0.80 & 0.80 \\
& F1 & 0.75 & 0.75 & 0.78 & \textbf{0.80} & 0.75 & \textbf{0.77} & 0.78 & 0.78 & 0.78 & 0.77 \\
\bottomrule
\end{tabular}
\end{adjustbox}
\label{tab:averaged_scores}
\end{table*}

\subsection{RQ2: To what extent do prompting
techniques (SCoT, k-shot) optimize user story generation?}

%When SCoT reasoning is used without any example-based prompting, as shown in table \ref{tab:averaged_scores}, its effect is dependent on model size.
%As shown in Table~\ref{tab:averaged_scores}, LLAMA-70B and LLAMA-405B achieve up to $0.04$ F1 score improvements in the 101-200 and 201-350 NLOC ranges when SCoT is used. 
%In contrast, LLAMA-8B, GPT-4o-mini, and fine-tuned GPT-4o-mini show either no performance changes or, in some cases, lower F1 scores across all NLOC ranges.

%This limited effectiveness in smaller models likely indicates an insufficient representational capacity to incorporate the additional reasoning instructions. These smaller models frequently ignore structured reasoning prompts or omit the thought process altogether. Conversely, larger LLAMA variants better leverage SCoT's control-flow cues, producing more complete user story formulations.

The impact of SCoT reasoning on model performance is contingent upon model size, as illustrated in Table~\ref{tab:averaged_scores}. Larger models, such as LLAMA-70B and LLAMA-405B, show improvements in F1 scores, up to $4\%$ in the 101-200 and 201-350 NLOC ranges when SCoT reasoning is employed. In contrast, smaller models, including LLAMA-8B, GPT-4o-mini, and fine-tuned GPT-4o-mini, exhibit either no significant performance change or a reduction in F1 scores across all NLOC ranges. This suggests that smaller models may lack the representational capacity to effectively integrate the additional reasoning instructions provided by SCoT. %These models often fail to utilize structured reasoning prompts or may omit the reasoning process entirely. On the other hand, larger LLAMA variants are better equipped to leverage SCoT's control-flow guidance, resulting in more comprehensive and accurate user story generation.

\begin{figure}[h]
\caption{Comparing all prompting techniques using LLAMA-8B model to base LLAMA-70B model.}
\centering
\includegraphics[width=\columnwidth]{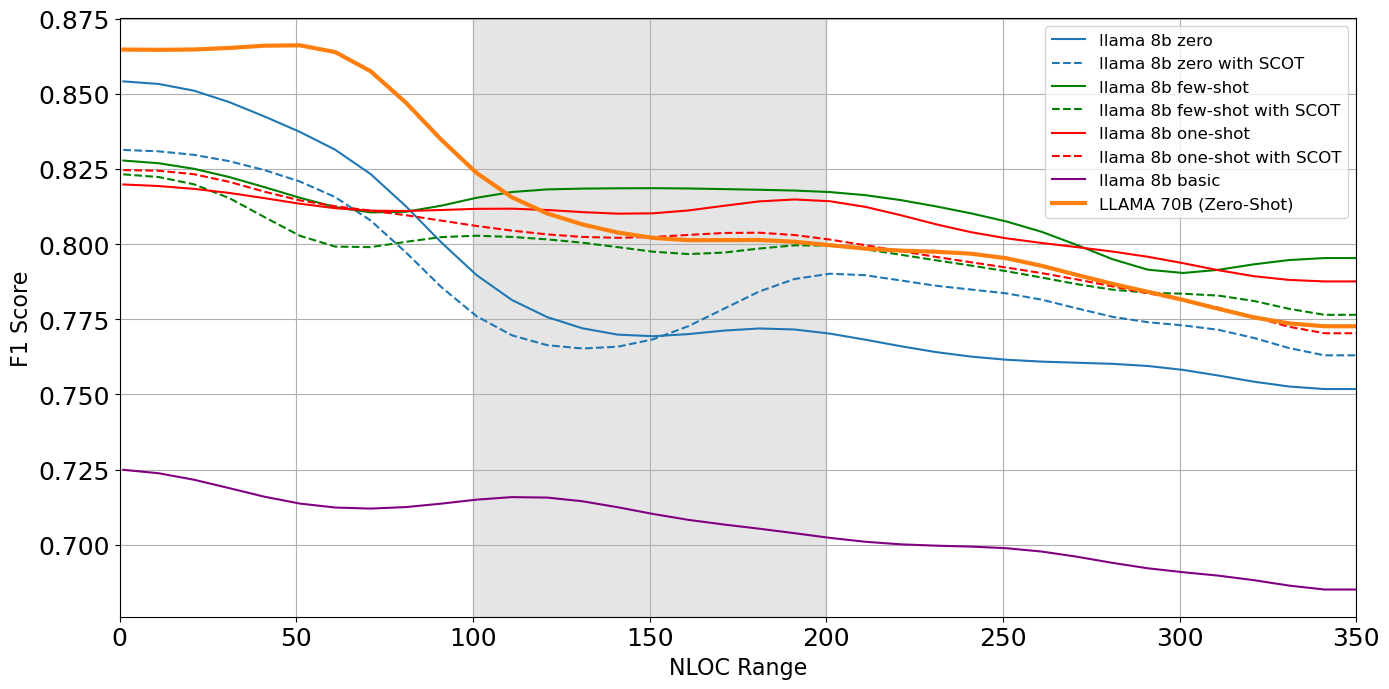}
\label{fig:f1 score for all prompt}
\end{figure}

As for the example-based prompting, our findings show that having one example (one-shot) is beneficial, while additional examples (few-shot) offer diminishing returns. 

Table~\ref{tab:averaged_scores_8b} shows that moving from zero-shot to one-shot prompt increases LLAMA-8B's F1 scores by $4\%$ in the 101-200 and 201-350 NLOC ranges. However, few-shot prompting adds only $1\%$ improvements over one-shot approaches. This pattern suggests that LLAMA-8B can effectively extract implicit rules from a single worked example. SCoT shows minimal impact when combined with example-based prompting. Table~\ref{tab:averaged_scores_8b} shows that adding SCoT to one-shot and few-shot configurations produces negligible F1 changes or sometimes reduces it.

\begin{table}[ht]
\centering
\caption{Averaged Precision, Recall, and F1 for LLAMA 8B across NLOC bands, SCoT usage, and prompt types. Bold = SCoT improvement; $\uparrow$ = One/Few-Shot No-SCoT gain over Zero-Shot No-SCoT.}
\begin{adjustbox}{max width=0.48\textwidth}
\begin{tabular}{cl|cc|cc|cc}
\toprule
\textbf{NLOC} &  & \multicolumn{2}{c|}{\textbf{Zero-Shot}} & \multicolumn{2}{c|}{\textbf{One-Shot}} & \multicolumn{2}{c}{\textbf{Few-Shot}} \\
\cmidrule(lr){3-4} \cmidrule(lr){5-6} \cmidrule(lr){7-8}
& & No SCoT & With SCoT & No SCoT & With SCoT & No SCoT & With SCoT \\
\midrule
\multirow{3}{*}{1--100}
& P  & 0.82 & 0.81 & 0.81 & 0.81 & 0.82 & 0.81 \\
& R  & 0.84 & 0.82 & 0.81 & 0.81 & 0.81 & 0.80 \\
& F1 & 0.83 & 0.81 & 0.81 & 0.81 & 0.81 & 0.80 \\
\midrule
\multirow{3}{*}{101--200}
& P  & 0.75 & 0.75 & 0.80$^{\uparrow}$ & 0.78 & 0.81$^{\uparrow}$ & 0.78 \\
& R  & 0.79 & 0.77 & 0.82$^{\uparrow}$ & 0.82 & 0.82$^{\uparrow}$ & 0.81 \\
& F1 & 0.77 & 0.77 & 0.81$^{\uparrow}$ & 0.80 & 0.81$^{\uparrow}$ & 0.79 \\
\midrule
\multirow{3}{*}{201-350}
& P  & 0.72 & 0.72 & 0.79$^{\uparrow}$ & 0.76 & 0.79$^{\uparrow}$ & 0.77 \\
& R  & 0.79 & 0.79 & 0.80$^{\uparrow}$ & \textbf{0.81} & 0.81$^{\uparrow}$ & 0.80 \\
& F1 & 0.75 & 0.75 & 0.79$^{\uparrow}$ & 0.78 & 0.80$^{\uparrow}$ & 0.78 \\
\bottomrule
\end{tabular}
\end{adjustbox}

\label{tab:averaged_scores_8b}
\end{table}

Prompt engineering improves LLAMA-8B performance, enabling it to match larger models despite parameter limitations. 

As shown in Figure~\ref{fig:f1 score for all prompt}, the basic LLAMA-8B configuration achieves a maximum F1 score of 0.73 across all complexity levels when using minimal prompting instructions. One-shot and few-shot prompting deliver performance improvements for LLAMA-8B, allowing it to consistently outperform LLAMA-70B after the 120 NLOC mark. SCoT showed mixed results, when combined with example-based prompting, it produces \(\pm 1\%\)  F1 change, and when used without exemplar prompting, it improved LLAMA-8B performance beyond the 150 NLOC mark. These findings highlight that strategic prompting can improve the performance of smaller models, especially for medium-complexity code generation tasks. Such tasks, which reflect typical functionality complexity in practice, provide a foundation for scaling this approach to larger codebases.

\begin{center}
\fbox{\parbox{0.94\linewidth}{%
\textbf{RQ2 Answer.} One-shot prompts yield the most substantial gains for LLAMA-8B, whereas few-shot and SCoT provide minimal additional benefit. Exemplar prompting alone allows the smaller 8B model to match or even exceed the performance of a 70B model for medium-sized code.}}
\end{center}

%\section{Discussion.}
%We compared the generated user stories with their ground-truth references, with samples available in the supplementary material \cite{suppdoc}. This comparison shows that for snippets up to 150 NLOCs the model preserves both the actor–goal and benefit clauses with minimal changes, differing mainly in lexical choice (e.g. \emph{digits required}→\emph{digits the sum will contain}). Beyond that NLOC range, the goal clause remains stable but the benefit clause begins to drift, often adopting more elaborate or technical wording and being more detailed (e.g. \emph{reduce the overall travel cost}→\emph{minimize the total travel distance and its associated financial expenditure}). Nevertheless, all generated stories convey the same intent as their references, confirming that the model’s paraphrasing stays semantically faithful.

\section{Threats to Validity}

Our study relies on CodeNet’s competitive programming solutions, which may not reflect the complexity of large-scale production systems. We mitigate this by noting that these tasks provide self-contained functional units suitable for proof-of-concept validation. The focus on snippets up to 350 NLOC also simplifies the problem space, though this range aligns with typical class and function sizes observed in popular GitHub repositories and we hierarchical extraction and chunking can be used as strategies for larger systems in future work. Our restriction to C++ limits generalizability, yet its popularity and prevalence in codebases justifies the choice, and the approach is transferable to other languages. Finally, while ground-truth user stories involve subjective interpretation, they were authored by three C++ proficient annotators with reconciled differences and achieved strong agreement (Cohen’s $\kappa$ = 0.86), supporting the reliability of our dataset.

\section{Conclusion and Future Work}
We analyzed 1,750 C++ snippets and demonstrated that large language models can successfully reverse-engineer user stories from source code with at least Adequate-paraphrase fidelity. A key finding is that a lightweight 8B model can match the performance of a much larger 70B model, enabling cost-effective and locally deployable solutions. Looking forward, we plan to extend our approach to larger codebases through hierarchical extraction, where function-level stories aggregate into class- and module-level narratives using chunking strategies for code exceeding 200 NLOC. Beyond scalability, we envision user story recovery as a validation mechanism for AI coding assistants. This verification can detect divergences from user intent, enabling automatic hallucination detection and self-correction. Such validation directly addresses a pressing need as AI tools become increasingly integrated into software development workflows.

\section{Acknowledgment}
This work was supported by NSERC Discovery Grant RGPIN-2024-06511.

\bibliographystyle{IEEEtran}
\bibliography{biblio}

\begin{thebibliography}{10}
\providecommand{\url}[1]{#1}
\csname url@samestyle\endcsname
\providecommand{\newblock}{\relax}
\providecommand{\bibinfo}[2]{#2}
\providecommand{\BIBentrySTDinterwordspacing}{\spaceskip=0pt\relax}
\providecommand{\BIBentryALTinterwordstretchfactor}{4}
\providecommand{\BIBentryALTinterwordspacing}{\spaceskip=\fontdimen2\font plus
\BIBentryALTinterwordstretchfactor\fontdimen3\font minus
  \fontdimen4\font\relax}
\providecommand{\BIBforeignlanguage}[2]{{%
\expandafter\ifx\csname l@#1\endcsname\relax
\typeout{** WARNING: IEEEtran.bst: No hyphenation pattern has been}%
\typeout{** loaded for the language `#1'. Using the pattern for}%
\typeout{** the default language instead.}%
\else
\language=\csname l@#1\endcsname
\fi
#2}}
\providecommand{\BIBdecl}{\relax}
\BIBdecl

\bibitem{lucassen2015forging}
G.~Lucassen, F.~Dalpiaz, J.~M.~E. Van Der~Werf, and S.~Brinkkemper, ``Forging
  high-quality user stories: towards a discipline for agile requirements,'' in
  \emph{2015 IEEE 23rd international requirements engineering conference
  (RE)}.\hskip 1em plus 0.5em minus 0.4em\relax IEEE, 2015, pp. 126--135.

\bibitem{ramesh2010agile}
B.~Ramesh, L.~Cao, and R.~Baskerville, ``Agile requirements engineering
  practices and challenges: an empirical study,'' \emph{Information Systems
  Journal}, vol.~20, no.~5, pp. 449--480, 2010.

\bibitem{rahman2024automated}
T.~Rahman and Y.~Zhu, ``Automated user story generation with test case
  specification using large language model,'' \emph{arXiv preprint
  arXiv:2404.01558}, 2024.

\bibitem{chinnaswamy2024user}
A.~Chinnaswamy, B.~Sabarish, and R.~Deepak~Menan, ``User story based automated
  test case generation using nlp,'' in \emph{International Conference on
  Computational Intelligence in Data Science}.\hskip 1em plus 0.5em minus
  0.4em\relax Springer, 2024, pp. 156--166.

\bibitem{guizani2021long}
M.~Guizani, A.~Chatterjee, B.~Trinkenreich, M.~E. May, G.~J. Noa-Guevara, L.~J.
  Russell, G.~G. Cuevas~Zambrano, D.~Izquierdo-Cortazar, I.~Steinmacher, M.~A.
  Gerosa \emph{et~al.}, ``The long road ahead: Ongoing challenges in
  contributing to large oss organizations and what to do,'' \emph{Proceedings
  of the ACM on Human-Computer Interaction}, vol.~5, no. CSCW2, pp. 1--30,
  2021.

\bibitem{guizani2022perceptions}
M.~Guizani, B.~Trinkenreich, A.~A. Castro-Guzman, I.~Steinmacher, M.~Gerosa,
  and A.~Sarma, ``Perceptions of the state of d\&i and d\&i initiative in the
  asf,'' in \emph{Proceedings of the 2022 ACM/IEEE 44th International
  Conference on Software Engineering: Software Engineering in Society}, 2022,
  pp. 130--142.

\bibitem{steinmacher2015systematic}
I.~Steinmacher, M.~A.~G. Silva, M.~A. Gerosa, and D.~F. Redmiles, ``A
  systematic literature review on the barriers faced by newcomers to open
  source software projects,'' \emph{Information and Software Technology},
  vol.~59, pp. 67--85, 2015.

\bibitem{genaid2012connecting}
A.~Genaid \emph{et~al.}, ``Connecting user stories and code for test
  development,'' in \emph{2012 third international workshop on recommendation
  systems for software engineering (rsse)}.\hskip 1em plus 0.5em minus
  0.4em\relax IEEE, 2012, pp. 33--37.

\bibitem{fischbach2020specmate}
J.~Fischbach, A.~Vogelsang, D.~Spies, A.~Wehrle, M.~Junker, and
  D.~Freudenstein, ``Specmate: Automated creation of test cases from acceptance
  criteria,'' in \emph{2020 IEEE 13th international conference on software
  testing, validation and verification (ICST)}.\hskip 1em plus 0.5em minus
  0.4em\relax IEEE, 2020, pp. 321--331.

\bibitem{elallaoui2015automatic}
M.~Elallaoui, K.~Nafil, and R.~Touahni, ``Automatic generation of uml sequence
  diagrams from user stories in scrum process,'' in \emph{2015 10th
  international conference on intelligent systems: theories and applications
  (SITA)}.\hskip 1em plus 0.5em minus 0.4em\relax IEEE, 2015, pp. 1--6.

\bibitem{elallaoui2018automatic}
------, ``Automatic transformation of user stories into uml use case diagrams
  using nlp techniques,'' \emph{Procedia computer science}, vol. 130, pp.
  42--49, 2018.

\bibitem{gilson2020generating}
F.~Gilson, M.~Galster, and F.~Georis, ``Generating use case scenarios from user
  stories,'' in \emph{Proceedings of the international conference on software
  and system processes}, 2020, pp. 31--40.

\bibitem{raharjana2021user}
I.~K. Raharjana, D.~Siahaan, and C.~Fatichah, ``User stories and natural
  language processing: A systematic literature review,'' \emph{IEEE access},
  vol.~9, pp. 53\,811--53\,826, 2021.

\bibitem{nistala2022towards}
P.~V. Nistala, A.~Rajbhoj, V.~Kulkarni, S.~Soni, K.~V. Nori, and R.~Reddy,
  ``Towards digitalization of requirements: generating context-sensitive user
  stories from diverse specifications,'' \emph{Automated Software Engineering},
  vol.~29, no.~1, p.~26, 2022.

\bibitem{colavito2024leveraging}
G.~Colavito, F.~Lanubile, N.~Novielli, and L.~Quaranta, ``Leveraging gpt-like
  llms to automate issue labeling,'' in \emph{2024 IEEE/ACM 21st International
  Conference on Mining Software Repositories (MSR)}.\hskip 1em plus 0.5em minus
  0.4em\relax IEEE, 2024, pp. 469--480.

\bibitem{abed2024ai}
O.~Abed, K.~Nebe, and A.~B. Abdellatif, ``Ai-generated user stories supporting
  human-centred development: An investigation on quality,'' in
  \emph{International Conference on Human-Computer Interaction}.\hskip 1em plus
  0.5em minus 0.4em\relax Springer, 2024, pp. 3--13.

\bibitem{marques2022enriching}
A.~B. Marques, A.~F. Costa, I.~Santos, and R.~Andrade, ``Enriching user stories
  with usability features in a remote agile project: a case study,'' in
  \emph{Proceedings of the XXI Brazilian Symposium on Software Quality}, 2022,
  pp. 1--10.

\bibitem{marques2024using}
N.~Marques, R.~R. Silva, and J.~Bernardino, ``Using chatgpt in software
  requirements engineering: A comprehensive review,'' \emph{Future Internet},
  vol.~16, no.~6, p. 180, 2024.

\bibitem{tikayat2023agile}
A.~Tikayat~Ray, B.~F. Cole, O.~J. Pinon~Fischer, A.~P. Bhat, R.~T. White, and
  D.~N. Mavris, ``Agile methodology for the standardization of engineering
  requirements using large language models,'' \emph{Systems}, vol.~11, no.~7,
  p. 352, 2023.

\bibitem{lee2024github}
J.~Y. Lee, S.~Kang, J.~Yoon, and S.~Yoo, ``The github recent bugs dataset for
  evaluating llm-based debugging applications,'' in \emph{2024 IEEE Conference
  on Software Testing, Verification and Validation (ICST)}.\hskip 1em plus
  0.5em minus 0.4em\relax IEEE, 2024, pp. 442--444.

\bibitem{macneil2024decoding}
S.~MacNeil, P.~Denny, A.~Tran, J.~Leinonen, S.~Bernstein, A.~Hellas, S.~Sarsa,
  and J.~Kim, ``Decoding logic errors: a comparative study on bug detection by
  students and large language models,'' in \emph{Proceedings of the 26th
  Australasian Computing Education Conference}, 2024, pp. 11--18.

\bibitem{wong2023natural}
M.-F. Wong, S.~Guo, C.-N. Hang, S.-W. Ho, and C.-W. Tan, ``Natural language
  generation and understanding of big code for ai-assisted programming: A
  review,'' \emph{Entropy}, vol.~25, no.~6, p. 888, 2023.

\bibitem{kim2012identifying}
M.~Kim, D.~Notkin, D.~Grossman, and G.~Wilson, ``Identifying and summarizing
  systematic code changes via rule inference,'' \emph{IEEE Transactions on
  Software Engineering}, vol.~39, no.~1, pp. 45--62, 2012.

\bibitem{zhang2022survey}
C.~Zhang, J.~Wang, Q.~Zhou, T.~Xu, K.~Tang, H.~Gui, and F.~Liu, ``A survey of
  automatic source code summarization,'' \emph{Symmetry}, vol.~14, no.~3, p.
  471, 2022.

\bibitem{bansal2021project}
A.~Bansal, S.~Haque, and C.~McMillan, ``Project-level encoding for neural
  source code summarization of subroutines,'' in \emph{2021 IEEE/ACM 29th
  International Conference on Program Comprehension (ICPC)}.\hskip 1em plus
  0.5em minus 0.4em\relax IEEE, 2021, pp. 253--264.

\bibitem{leclair2021ensemble}
A.~LeClair, A.~Bansal, and C.~McMillan, ``Ensemble models for neural source
  code summarization of subroutines,'' in \emph{2021 IEEE International
  Conference on Software Maintenance and Evolution (ICSME)}.\hskip 1em plus
  0.5em minus 0.4em\relax IEEE, 2021, pp. 286--297.

\bibitem{sun2024source}
W.~Sun, Y.~Miao, Y.~Li, H.~Zhang, C.~Fang, Y.~Liu, G.~Deng, Y.~Liu, and
  Z.~Chen, ``Source code summarization in the era of large language models,''
  \emph{arXiv preprint arXiv:2407.07959}, 2024.

\bibitem{yun2024project}
S.~Yun, S.~Lin, X.~Gu, and B.~Shen, ``Project-specific code summarization with
  in-context learning,'' \emph{Journal of Systems and Software}, vol. 216, p.
  112149, 2024.

\bibitem{zhu2024effectiveness}
J.~Zhu, Y.~Miao, T.~Xu, J.~Zhu, and X.~Sun, ``On the effectiveness of large
  language models in statement-level code summarization,'' in \emph{2024 IEEE
  24th International Conference on Software Quality, Reliability and Security
  (QRS)}.\hskip 1em plus 0.5em minus 0.4em\relax IEEE, 2024, pp. 216--227.

\bibitem{octoverse2022}
{GitHub}, ``The top programming languages on github — octoverse 2022,''
  \url{https://octoverse.github.com/2022/top-programming-languages}, 2022, c++
  ranked 6th with ~8\% of contributions; accessed 30 May 2025.

\bibitem{tiobe2024}
{TIOBE Software}, ``Tiobe index for 2024,''
  \url{https://www.tiobe.com/tiobe-index/}, 2024, c++ ranked in top 3
  throughout 2024; accessed 31 May 2025.

\bibitem{puri2021codenet}
R.~Puri, D.~S. Kung, G.~Domeniconi, and et~al., ``Project codenet: A
  large-scale ai for code dataset for learning a diversity of coding tasks,''
  in \emph{NeurIPS Datasets \& Benchmarks}, 2021, dataset overview and
  benchmark definitions.

\bibitem{ibm_codenet_2021}
{IBM Research}, ``Project codenet – ibm developer exchange,''
  \url{https://developer.ibm.com/exchanges/data/all/project-codenet/}, 2021.

\bibitem{liu2023lost}
N.~F. Liu, N.~Scharli, A.~Lewkowycz, J.~P. Lee, K.~Guu, I.~Schlag, A.~M. Dai,
  A.~W. Yu, M.~van Zuylen, and D.~Zhou, ``Lost in the middle: How language
  models use long contexts,'' \emph{arXiv preprint arXiv:2307.03172}, 2023.

\bibitem{cohen1960coefficient}
J.~Cohen, ``A coefficient of agreement for nominal scales,'' \emph{Educational
  and psychological measurement}, vol.~20, no.~1, pp. 37--46, 1960.

\bibitem{suppdoc}
\BIBentryALTinterwordspacing
Anonymized, ``Supplemental material for reverse engineering user stories from
  code using large language models,'' 2025, supplemental Material. [Online].
  Available: \url{https://doi.org/10.5281/zenodo.15849709}
\BIBentrySTDinterwordspacing

\bibitem{meta8bprice}
{Meta AI}, ``meta-llama-3-8b – run with an api on replicate,'' 2024.

\bibitem{deepseekprice}
{DeepSeek}, ``Models \& pricing – deepseek api docs,'' 2024.

\bibitem{openaiapi}
{OpenAI}, ``Api pricing,'' \url{https://openai.com/api/pricing/}, 2025,
  gPT-4o-mini price card, accessed 30 May 2025.

\bibitem{fujita2024llm}
M.~Fujita, T.~Onaga, and Y.~Kano, ``Llm tuning and interpretable cot: Kis team
  in coliee 2024,'' in \emph{JSAI International Symposium on Artificial
  Intelligence}.\hskip 1em plus 0.5em minus 0.4em\relax Springer, 2024, pp.
  140--155.

\bibitem{li2023structured}
J.~Li, G.~Li, Y.~Li, and Z.~Jin, ``Structured chain-of-thought prompting for
  code generation,'' \emph{ACM Transactions on Software Engineering and
  Methodology}, 2023.

\bibitem{replicate2024}
\BIBentryALTinterwordspacing
Replicate - run machine learning models in the cloud. Accessed: 2024-12-05.
  [Online]. Available: \url{https://replicate.com/}
\BIBentrySTDinterwordspacing

\bibitem{deepseek2024r1}
DeepSeek-AI, ``Deepseek-r1: Incentivizing reasoning capability in llms via
  reinforcement learning,'' 2024.

\end{thebibliography}

\end{document}